\documentclass[aps, superscriptaddress, floatfix, 12pt, nofootinbib]{revtex4}
\usepackage{graphicx}
\usepackage{amsmath}
\usepackage{amssymb}
\usepackage{color}

\definecolor{blue}{rgb}{0,0,1}

\newcommand{\be}{\begin{eqnarray}}
\newcommand{\ee}{\end{eqnarray}}

\newcommand{\Tr}{\mathrm{Tr}}
\newcommand{\E}{\mathrm{e}}
\newcommand{\nn}{\nonumber }
\def\slash#1{\setbox0=\hbox{$#1$}               
   \dimen0=\wd0                                 
   \setbox1=\hbox{/} \dimen1=\wd1               
   \ifdim\dimen0>\dimen1                        
      \rlap{\hbox to \dimen0{\hfil/\hfil}}      
      #1                                        
   \else                                        
      \rlap{\hbox to \dimen1{\hfil$#1$\hfil}}   
      /                                         
   \fi}                                         %

\begin{document}

\title{Volume and Quark Mass Dependence\\ of the Chiral Phase Transition}
\author{J. Braun}
\affiliation{Institute for Theoretical Physics, University of
  Heidelberg, Philosophenweg 19, 69120 Heidelberg, Germany}
\author{B. Klein}
\affiliation{GSI, Planckstrasse 1, 64159 Darmstadt, Germany}
\author{H.-J. Pirner}
\affiliation{Institute for Theoretical Physics, University of
  Heidelberg, Philosophenweg 19, 69120 Heidelberg, Germany}
\affiliation{Max-Planck-Institut f\"ur Kernphysik, Saupfercheckweg 1,
  69117 Heidelberg, Germany}
\author{A. H. Rezaeian}
\affiliation{Institute for Theoretical Physics, University of
  Heidelberg, Philosophenweg 19, 69120 Heidelberg, Germany}

\date{\today}

\begin{abstract}
We investigate chiral symmetry restoration in finite spatial volume
and at finite temperature by calculating the dependence of the chiral
phase transition temperature $T_{c}$ on the size of the 
spatial volume and the current-quark mass for the quark-meson model, using 
the proper-time Renormalization Group approach. We find that the 
critical temperature is weakly dependent on the size of the 
spatial volume for large current-quark
masses, but depends strongly on it for small current-quark masses.
In addition, for small volumes we observe a dependence on the choice
of quark boundary conditions. 
\end{abstract}

\maketitle

\section{Introduction}
\label{sec:intro}
Phase transitions in Quantum Chromodynamics (QCD) are currently
very actively researched. Most of the attention is focused
on the phase transition at finite baryon density and temperature
\cite{Fodor:2001pe, 
  Fodor:2004nz, Allton:2002zi, Allton:2003vx, Philipsen:2005mj, 
deForcrand:2002ci, deForcrand:2003hx, Schaefer:2004en}, where 
the existence of a critical point in the phase diagram is not yet
conclusively settled \cite{Gavai:2004sd, Allton:2003vx,
  Allton:2005gk}. 
Even at vanishing density, the order of the phase transition is
still under discussion 
\cite{Pisarski:1983ms, D'Elia:2005bv, Karsch:2000kv, Karsch:2001nf}. 

While QCD is perturbative at large momentum scales, 
the low-energy limit of the theory is dominated by non-perturbative phenomena.
This makes non-perturbative methods indispensable, in particular for
investigations of the phase transition. Effective
low-energy theories such 
as chiral perturbation theory \cite{Weinberg:1978kz, Gasser:1983yg,
  Gasser:1984gg} describe
the low-energy limit of QCD well, but cannot address the restoration
of chiral symmetry and the deconfinement transition: 
These questions require a connection to the high-momentum degrees of freedom.
Lattice gauge theory, on the other hand, yields non-perturbative results,
allows an investigation of the phase transition, and can in principle
provide the necessary effective couplings for the description by a
low-energy effective theory \cite{Farchioni:2003nf,
  Farchioni:2004fs}. But even in the light of recent 
advances with light fermions on the lattice \cite{Chen:2000zu, Aoki:2004ht},
most lattice simulations still require extrapolations to small,
realistic pion masses. In any case extrapolations to the continuum
limit and to infinite volume are necessary. In particular, finite
volume effects are more severe when the pion mass
approaches the chiral limit \cite{Colangelo:2002hy, Colangelo:2003hf,
  Colangelo:2005gd, Braun:2004yk, Braun:2005gy}.
Therefore, the influence of the finite volume for small pion masses
should be studied with other methods in parallel.

In addition, lattice gauge theory provides little guidance to understand
the emergence of low-energy dynamics.
The interplay between lattice gauge theory and other non-perturbative
methods such as Dyson-Schwinger equations \cite{Fischer:2002hn,
  Maas:2004se, Maas:2005hs, Maas:2005ym, Fischer:2005nf} and the
Renormalization Group 
\cite{Pawlowski:2003hq, Gies:2002af, Fischer:2004uk, Gies:2005as,
  Braun:2005uj} should prove fruitful to 
further our understanding. There is also a need
for model systems that describe particular aspects of the dynamical
generation of the low-momentum physics from the high-momentum theory.
One example is the description of chiral symmetry breaking via the
Nambu--Jona-Lasinio model \cite{Nambu:1961tp} and its
modifications. The quark-meson model that we use in the present 
work belongs to this class of models \cite{Bijnens:1995ww}.

Of course, such a model approach cannot answer questions outside the
applicability of the model. For example, the order of the phase transition
is in our case already fixed by the $O(4)$-symmetry of the model,
while the order of the transition in QCD has not yet been unambiguously
determined \cite{D'Elia:2005bv, Philipsen:2005mj}.  
We must also limit our investigation to the chiral phase transition. In
QCD, there is no requirement that the chiral
and deconfinement phase transitions occur at the same point
\cite{Pisarski:1983ms}, although 
so far there is no indication for two transitions. 
On the other hand, the model has been successfully
used to investigate the quark mass dependence of the chiral transition
\cite{Berges:1997eu} and the critical behavior at finite density
\cite{Schaefer:2004en} with Renormalization Group methods. 
It has also recently been combined with Polyakov loop results from the
lattice to describe thermodynamical observables from lattice QCD
\cite{Ratti:2005jh}. 

The Renormalization Group (RG) is an important tool for the
investigation of non-perturbative physics \cite{Wilson:1973jj, Wegner:1972ih,
  Wetterich:1992yh, Liao:1994fp}. 
In particular, RG methods are well suited to describe physics
across different momentum scales, and generation of the low-energy
effective theories from the dynamics. While we do not directly address this
issue here, the study of critical behavior is of course also well
within the scope of an RG approach, and in the present context it
has been applied to determine critical exponents, for example for the
quark-meson model \cite{Berges:1997eu, Schaefer:1999em}.  

In this paper, we consider the chiral phase transition in the
framework of the quark-meson model. We will apply Renormalization
Group methods to calculate the transition temperature, its dependence
on the quark mass, and its dependence on the size of a finite volume.
In addition, we investigate the effects of different boundary
conditions for the quark fields. In calculations based on an
effective field theory like chiral perturbation theory, chiral
symmetry breaking is assumed from the beginning and the values of the effective
low-energy constants are fixed. In contrast, in our model chiral
symmetry is broken dynamically and effects of the finite volume on
quark condensation are taken into account. We believe that such
effects could still be important in simulations at the current lattice
sizes of order $L \simeq 2 \; \mathrm{fm}$, in particular for
realistic quark masses.

In finite volume, strictly speaking no phase transition is possible,
since non-analyticities cannot appear in the thermodynamic potential (see
e.g. \cite{Goldenfeld:1992qy}). In general, the investigation
of phase transitions and critical behavior from results obtained in a
finite volume is difficult and requires an extrapolation to the
large-volume limit. In addition, if a symmetry is restored across the
transition, this usually requires the introduction of an external
field which explicitly breaks the symmetry.
Even if there is no true order parameter
that vanishes strictly in one of the phases, rapid changes over a
small temperature range are an indication of a (crossover)
transition. Often, peaks in susceptibilities or other
higher-order derivatives of the thermodynamic potential are used as
criterion to define a pseudo-transition. Here we propose to use the
mass of the scalar mode, which corresponds to the inverse correlation
length for fluctuations in the quark condensate, to
identify the transition point: A distinct minimum of the mass appears at
almost the same temperature at which the chiral quark condensate
drops rapidly. We stress that the implementation of an explicitly
chiral-symmetry-breaking 
term is essential, since the finite-volume system will otherwise always
be in a regime with restored chiral symmetry, once all quantum fluctuations
are taken into account.

Lattice simulations are affected by similar problems. Thus, finite 
volume effects are actually of profound importance in lattice
determinations of the order of the phase transition. 
For the determination of the universality class and the critical
exponents of the
transition, a scaling analysis of thermodynamic observables is
necessary \cite{Karsch:1994hm, Aoki:1998wg, Bernard:1999fv, D'Elia:2005bv,
  Philipsen:2005mj}. It remains difficult to assess whether current lattice
sizes are sufficiently large to observe the expected scaling
\cite{Aoki:1998wg, D'Elia:2005bv, Philipsen:2005mj}. On the other
hand, the volume dependence of the (pseudo-) critical behavior can be
turned into a tool for the analysis: Finite-size scaling of the
results has been 
used to test the compatibility of critical exponents with
lattice data \cite{Aoki:1998wg, D'Elia:2005bv}.
We expect that future progress in RG analysis will provide much useful
insight into these questions. 

The paper is divided into five sections. In the next section, we will
give a short review of the quark-meson model and the derivation of
RG flow equations in the proper-time formulation of the RG. In section
\ref{sec:cpt_iv}, we will concentrate on the chiral phase transition
in infinite volume, and its dependence on the parameter of explicit
chiral symmetry breaking. 
In section \ref{sec:cpt_fv}, we will then look at the finite volume
effects and the influence of the additional momentum scale introduced
by the finite volume. We close with a summary and conclusions in section
\ref{sec:conclusions}.

\section{RG-Flow Equations for the Quark-Meson Model}
\label{sec:flow-equations}
To determine the chiral phase transition
temperature for finite volumes and finite current quark masses, we
use the chiral quark-meson model\footnote{In the present approach to
  the phase transition at vanishing baryon density, we
  do not include vector mesons. The role of the vector
meson in medium is not yet completely understood \cite{Asakawa:1989bq,
Lutz:1992dv,Boyd:1994np,Rezaeian:2005nm, Rezaeian:2005zy}. An analysis on
the lattice suggests that at high temperature the vector coupling is
small compared to the scalar coupling \cite{Boyd:1994np}.}. 
This model is an $O(4)$-invariant linear
$\sigma$-model with $N_{f}^{2}=4$ mesonic degrees of freedom
$(\sigma,\vec{\pi})$ 
coupled to $N_{f}=2$ flavors of constituent quarks in an
$SU(2)_{L}\times SU(2)_{R}$ 
invariant way. It is an effective low-energy
model for dynamical spontaneous chiral symmetry breaking at intermediate
scales of $k\lesssim\Lambda_{UV}\approx1.5\;\mathrm{GeV}$, but it 
does not contain gluonic degrees of freedom
and is not confining. The ultraviolet (UV)
scale $\Lambda_{UV}\approx1.5\;\mathrm{GeV}$ is determined by the
validity of a hadronic representation of QCD. At the scale $\Lambda_{UV}$,
the quark-meson model is defined by the bare effective action 
\begin{equation}
\Gamma_{\Lambda_{UV}}[\phi]=
\int d^{4}x \left\{
\bar{q}(\slash{\partial}+gm_{c})q +
g\bar{q}(\sigma+i\vec{\tau}\cdot\vec{\pi}\gamma_{5})q 
+\frac{1}{2}(\partial_{\mu}\phi)^{2}+U_{\Lambda_{UV}}(\phi)\right\} 
\end{equation}
with a current quark mass term $gm_{c}$ which explicitly breaks
the chiral symmetry, and with
$\phi^{\mathrm{{T}}}=(\sigma,\pi^{1},\pi^{2},\pi^{2})$. 
The mesonic potential is characterized by two couplings: 
\begin{equation}
U_{\Lambda_{UV}}(\phi) =
\frac{1}{2}m_{UV}^{2}\phi^{2}+\frac{1}{4}\lambda_{UV}(\phi^{2})^{2}
\,.\label{eq:pot_UV} 
\end{equation}
In a Gaussian approximation, we can perform the functional integration
of the bosonic and fermionic fields and obtain the one-loop effective
action for the scalar fields $\phi$,
\begin{equation}
\Gamma[\phi] = \Gamma_{\Lambda_{UV}}[\phi] -
\Tr\log\left(\Gamma_{F}^{(2)}[\phi]\right) +
\frac{1}{2}\Tr\log\left(\Gamma_{B}^{(2)}[\phi]\right) \label{eq:1loop}
\end{equation}
where $\Gamma_{B}^{(2)}[\phi]$ and $\Gamma_{F}^{(2)}[\phi]$ are the 
inverse two-point functions.
We neglect a possible space dependence of the expectation
value and take the wave-function renormalization and the Yukawa-coupling
to be constant. Since the traces in Eq.~\eqref{eq:1loop} are infrared
(IR) divergent, we use the Schwinger proper-time representation of
the logarithms and introduce an infrared cutoff
function\footnote{Although the RG flow equations themselves depend on
  the particular form of the cutoff function, physical quantities
  calculated from the RG flow should not depend on the choice of the
  cutoff function in the  
limit $k \rightarrow 0$.} 
$f_{a} (\tau k^{2})$, where the variable
$\tau$ denotes Schwinger's proper time and $k$ is a cutoff scale. 
The derivative of the cutoff function with respect to the scale $k$ is
given by 
\begin{equation}
k\frac{\partial}{\partial k}f_{a}(\tau k^{2}) =
-\frac{2}{\Gamma(a+1)}(\tau k^{2})^{a+1}e^{-\tau k^{2}} \,.
\label{eq:cutoff-fct}
\end{equation}
The inverse two-point functions in Eq.~\eqref{eq:1loop} depend on the
second derivatives of the effective 
potential $U$. By replacing the bare masses and couplings in the
inverse two-point functions with the scale-dependent quantities, we
obtain the so-called renormalization group improved flow equation
for the effective potential $U_{k}$, in infinite volume for zero temperature
and finite current quark mass \cite{Braun:2004yk}:
\begin{eqnarray}
k\frac{\partial}{\partial k} U_{k}(\sigma,\vec{\pi}^{2},T &
 \rightarrow & 0,L\to\infty) = {\displaystyle
 \frac{k^{2(a+1)}}{16a(a-1)\pi^{2}} \Bigg\{ -
 \frac{4N_{c}N_{f}}{(k^{2}+M_{q}^{2}(\sigma,\vec{\pi}^{2}))^{a-1}}}
 \nonumber \\
 &  &  + \frac{1}{(k^{2}+M_{\sigma}^{2}(\sigma,\vec{\pi}^{2}))^{a-1}} 
+ \frac{N_{f}^{2}-1}{(k^{2}+M_{\pi}^{2}(\sigma,\vec{\pi}^{2}))^{a-1}}
 \Bigg\}\,. \label{eq:IV_fe}
\end{eqnarray}
Integrating the flow equation from the UV scale to $k \to 0$, we
obtain an effective potential in which quantum corrections from all
scales have been systematically included.

Since we allow for explicit symmetry breaking, the $O(4)$-symmetry
of the effective potential is lost. However, it remains $O(3)$-symmetric
in the pion-subspace, so that the pion-fields can only appear in the
combination $\vec{\pi}^{2}$ on the right-hand side. In order to be
able to perform the Schwinger proper-time integration in infinite
volume, we have to choose $a\geq 2$ \cite{Schaefer:1999em}.

The meson masses $M_{\sigma}$ and $M_{\pi}$ in Eq.~\eqref{eq:IV_fe} 
are the eigenvalues of the
second-derivative matrix of the mesonic potential, cf. \cite{Braun:2004yk}
for an explicit representation, and the constituent quark mass $M_{q}$ 
is given by
\begin{equation}
M_{q}^{2}=g^{2}[(\sigma+m_{c})^{2}+\vec{\pi}^{2}]\,.\label{eq:mq}
\end{equation}
We generalize the renormalization group flow equations to a finite
four-dimensional Euclidean volume $L^{3}\times T$ by replacing the integrals
over the momenta in the evaluation of the trace in Eq.~\eqref{eq:1loop}
by a sum
\be
\int dp_{i}\,\ldots \rightarrow \frac{2\pi}{L} \sum_{n_{i} =
  -\infty}^{\infty}\ldots\,. 
\ee
The boundary conditions in the Euclidean time direction are fixed
by the statistics of the fields. The thermal Matsubara frequencies
take the values 
\be
\omega_{n_{0}}=2\pi n_{0}T\quad \mathrm{and} \quad \nu_{n_{0}} =
(2n_{0}+1) \pi T\,,
\ee
for bosons and for fermions, respectively, where the temperature is
denoted by $T$. However, we are free in the choice of boundary conditions
for the bosons and fermions in the space directions. In the following
we use the short-hand notation
\begin{equation}
p_{p}^{2}=\frac{4\pi^{2}}{L^{2}}\sum_{i=1}^{3}n_{i}^{2}\quad
\mathrm{and} \quad
p_{ap}^{2}=\frac{4\pi^{2}}{L^{2}}
\sum_{i=1}^{3}\left(n_{i}+\frac{1}{2}\right)^{2} 
\label{eq:fv_mom}
\end{equation}
for the three-momenta in the case of periodic (p) and anti-periodic
(ap) boundary conditions. We will consider both choices for the quark
fields, but employ only periodic boundary conditions for mesonic fields.
Then the flow equation for finite temperature and finite volume reads
\begin{eqnarray}
k\frac{\partial}{\partial k}U_{k}(\sigma,\vec{\pi}^{2},T,L) & = &
 \frac{k^{2(a+1)}}{\Gamma(a+1)} \frac{T}{L^{3}} {\displaystyle
 \sum_{n_{0}} \sum_{\vec{n}} \int_{0}^{\infty}d\tau \tau^{a} \Bigg(-4
 N_{c} N_{f}
 \E^{-\tau(k^{2} +\nu_{n_{0}}^{2} +p_{ap,p}^{2}
 +M_{q}^{2}(\sigma,\vec{\pi}^{2}))}}\nonumber \\ 
 &  & \qquad\qquad+\sum_{i=1}^{N_{f}^{2}=4} \E^{-\tau(k^{2}+
 \omega_{n_{0}}^{2}
 +p_{p}^{2}+M_{i}^{2}(\sigma,\vec{\pi}^{2}))}\Bigg).\,\label{eq:FV_fe_old} 
\end{eqnarray}
The sums in Eq.~\eqref{eq:FV_fe_old} run from $-\infty$ to $+\infty$,
where the vector $\vec{n}$ denotes $(n_{1},n_{2},n_{3})$. Since
we have to solve the flow equation numerically, we
rewrite it in terms of Jacobi-Elliptic-Theta functions: 
\begin{eqnarray}
k\frac{\partial}{\partial k}U_{k}(\sigma,\vec{\pi}^{2},T,L) & = &
 \frac{T}{L^{3}} \frac{(kL)^{2(a+1)}}{(4\pi)^{a+1}\Gamma(a+1)}
 {\displaystyle \Bigg( \!-4 N_{c} N_{f}
 \Theta_{ap,p}^{(F)}\Big(a,(k^{2}\!+\!
 M_{q}^{2}(\sigma,\vec{\pi}^{2}))L^{2},TL\Big)} \nonumber \\
 &  & \qquad \qquad + \sum_{i=1}^{N_{f}^{2}=4} \Theta_{p}^{(B)}
 \Big(a,(k^{2}\!+\!
 M_{i}^{2}(\sigma,\vec{\pi}^{2}))L^{2},TL\Big)\Bigg). \,\label{eq:FV_fe}
\end{eqnarray}
We have introduced the auxiliary (dimensionless) functions 
\be
\Theta_{ap}^{(F)}(a,\omega,t)&=& \int_{0}^{\infty}ds\,
s^{a}\E^{-\frac{s\omega}{4\pi}}
\vartheta_{ap}(st^{2})\Big(\vartheta_{ap}(s)\Big)^{3}\,,\label{eq:ThetaF_ap}\\ 
\Theta_{p}^{(F)}(a,\omega,t)&=&\int_{0}^{\infty}ds\,
s^{a}\E^{-\frac{s\omega}{4\pi}} \vartheta_{ap}(st^{2})
\Big(\vartheta_{p}(s)\Big)^{3}\,,\label{eq:ThetaF_p}\\ 
\Theta_{p}^{(B)}(a,\omega,t)&=& \int_{0}^{\infty}ds\,
s^{a}\E^{-\frac{s\omega}{4\pi}}
\vartheta_{p}(st^{2})\Big(\vartheta_{p}(s)\Big)^{3}\,,\label{eq:ThetaB} 
\ee
where $\vartheta_{p}$ and $\vartheta_{ap}$ are Jacobi-Elliptic-Theta
functions defined as
\begin{eqnarray}
\vartheta_{ap}(x) &=& \sum_{n=-\infty}^{\infty}
    \E^{-x\pi(n+\frac{1}{2})^{2}} = x^{-\frac{1}{2}} + 2
    \sum_{q=1}^{\infty}(-1)^{q} x^{-\frac{1}{2}} \E^{-\frac{\pi
    q^{2}}{x}}\,,\label{eq:tetha_ap} \\
\vartheta_{p}(x) &=& \sum_{n=-\infty}^{\infty} \E^{-x\pi n^{2}} =
    x^{-\frac{1}{2}} + 2\sum_{q=1}^{\infty} x^{-\frac{1}{2}}
    \E^{-\frac{\pi q^{2}}{x}}\,.\label{eq:theta_p}  
\end{eqnarray}
The representation in terms of these functions accelerates the
numerical calculations by 
a factor of about
a hundred, compared to the representation used in
Refs.~\cite{Braun:2004yk, Braun:2005gy}. 
The first representation in Eq.~\eqref{eq:tetha_ap} and \eqref{eq:theta_p}
is the standard Matsubara summation of the momenta. The second representation
on the right hand side in Eq.~\eqref{eq:tetha_ap} and
\eqref{eq:theta_p} is obtained by 
applying Poisson's formula to the first representation. One can use
this representation to separate the zero-temperature and
infinite-volume contribution of the flow equation. Indeed, using the
approximation $\vartheta_{ap}(s)=\vartheta_{p}(s)\approx s^{-\frac{1}{2}}$ in
Eqs.~\eqref{eq:ThetaF_ap}, \eqref{eq:ThetaF_p}, and \eqref{eq:ThetaB} yields
\begin{equation}
\Theta_{ap}^{(F)}(a,\omega,t) = \Theta_{p}^{(F)}(a,\omega,t) =
\Theta_{p}^{(B)}(a,\omega,t) \approx
\frac{1}{t}\frac{\Gamma(a-1)}{\omega^{a-1}}\,.
\end{equation}
Inserting this in Eq.~\eqref{eq:FV_fe}, we obtain the flow equation
\eqref{eq:IV_fe} for infinite volume and zero temperature. From now
on, we use $a=2$ for both the infinite-volume and finite-volume
calculations.

The flow equations \eqref{eq:IV_fe} and \eqref{eq:FV_fe} are partial
differential equations which can be solved by using a projection of these
flow equations on the following ansatz for the mesonic potential
\cite{Braun:2004yk,Braun:2005gy}: 
\begin{equation}
U_{k}(\sigma,\vec{\pi}^{2}) =
\sum_{i=0}^{N_{\sigma}}\sum_{j=0}^{i+j\le N_{\sigma}}
a_{ij}(k) (\sigma-\sigma_{0}(k))^{i}
(\sigma^{2}+\vec{\pi}^{2}-\sigma_{0}^{2}(k))^{j} 
\label{eq:pot_ansatz}
\end{equation}
Such a projection results in an infinite set of coupled first-order
differential equations for the coefficients $a_{ij}(k)$. In order
to solve this set of equations, we limit the sum in Eq.~\eqref{eq:pot_ansatz}
by choosing $N_{\sigma}=2$ \cite{Braun:2004yk, Braun:2005gy}. The
boundary conditions for the differential 
equations for the coefficients $a_{ij}(k)$ are determined at the
ultraviolet scale $k=\Lambda_{UV}$. For a given current quark mass $gm_{c}$,
we determine the initial conditions $a_{ij}(\Lambda_{UV})$ for
infinite four-dimensional Euclidean volume in such a way that we obtain
values of $m_{\pi}$ and $f_{\pi}=\sigma_{0}$ which are consistent
with chiral perturbation theory \cite{Colangelo:2003hf}. 
Consequently, our calculation cannot predict the values of $m_{\pi}$
and $f_{\pi}=\sigma_{0}$ in infinite volume, 
but allows to study the behavior of the masses and the pion decay
constant in a finite four-dimensional Euclidean volume.
The parameter sets which we have used for the calculations in
Sec.~\ref{sec:cpt_iv} and \ref{sec:cpt_fv} are listed in
Appendix~\ref{sec:Starting-values}, where the relations 
between the meson masses and the coefficients $a_{ij}$ are summarized
as well.

\section{Chiral Phase Transition Temperature in Infinite Volume}
\label{sec:cpt_iv}
\begin{figure}
\includegraphics[clip,scale=0.9]{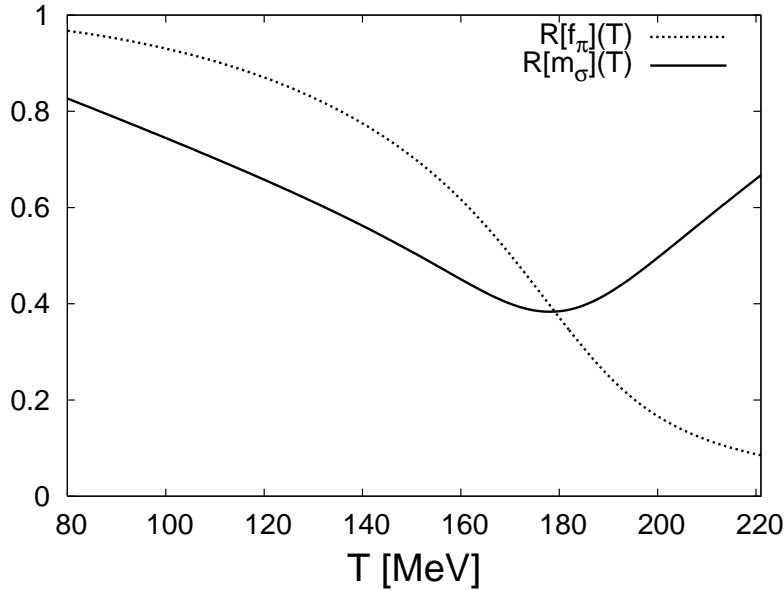}
\caption{Normalized sigma mass
  $R[m_{\sigma}](T)=\frac{m_{\sigma}(T)}{m_{\sigma}(0)}$ 
and normalized pion decay constant
  $R[f_{\pi}](T)=\frac{f_{\pi}(T)}{f_{\pi}(0)}$ as  
a function of the temperature, in infinite volume for a pion mass
  $m_{\pi}^{(0)}=100\,\mathrm{MeV}$.}
\label{cap:msig_fpi} 
\end{figure}

In this section, we discuss the dependence of the chiral phase transition
temperature in infinite volume on the zero-temperature pion mass
$m_{\pi}^{(0)}=m_{\pi}(T=0)$. 
In order to define a chiral phase transition temperature in the
presence
of explicit symmetry breaking, we use the dependence of the $\sigma$-mass
on the temperature. We define the phase transition
temperature $T_{c}$ through the minimum of the $\sigma$-mass,
\begin{equation}
\left.\frac{\partial m_{\sigma}(T)}{\partial
  T}\right|_{T=T_{c}} = 0 \quad \mathrm{{and} \quad}
  \left.\frac{\partial^{2}m_{\sigma}(T)}{\partial
  T^{2}}\right|_{T=T_{c}}>0\,.
\end{equation}
Alternatively, one can define the phase transition temperature
as turning point of the pion decay constant as a function of
temperature\footnote{There is  
no unique definition for the crossover or the pseudo-critical temperature. 
For example, the critical temperature $T_c$ can also be
defined as the temperature at which $f_{\pi}$ reaches half its
zero-temperature value \cite{Klevansky:1992qe}. The results for $T_c$
obtained with such a definition will in essence agree with our
results, as suggested  by Fig.~\ref{cap:msig_fpi}.}.
We have checked that the values for the chiral phase
transition temperature $T_{c}$ obtained from these two different definitions
agree within a few percent. For example, in Fig.~\ref{cap:msig_fpi}
we compare the normalized $\sigma$-mass $R[m_{\sigma}](T) =
\frac{m_{\sigma}(T)}{m_{\sigma}(0)}$ and the normalized pion decay constant
$R[f_{\pi}](T)=\frac{f_{\pi}(T)}{f_{\pi}(0)}$ for
$m_{\pi}^{(0)}=100\,\mathrm{{MeV}}$. 
One observes that both definitions for the critical temperature yield
for practical purposes the same result.

\begin{figure}
\includegraphics[clip,scale=0.9]{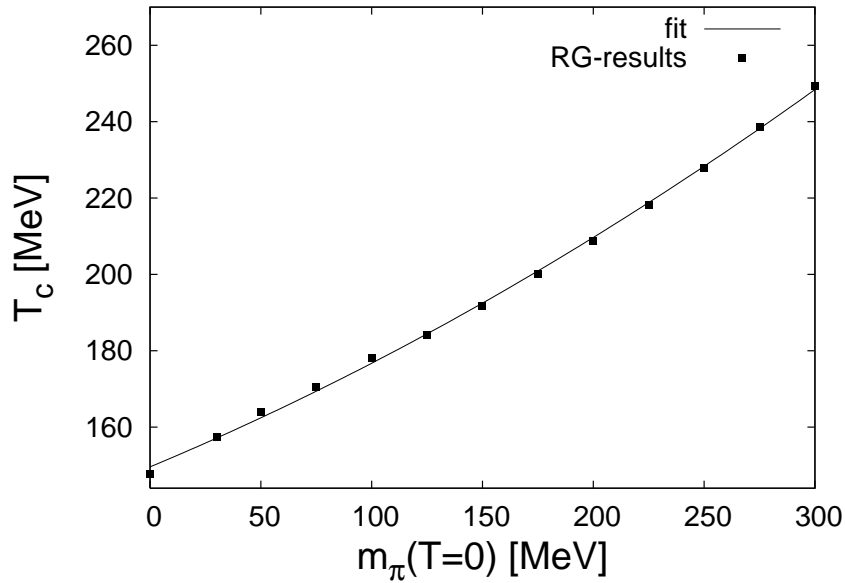}
\caption{\label{cap:Chiral-phase-transition}Chiral phase transition temperature $T_{c}$ in infinite volume as a function of $m_{\pi}(T=0)$.
The dots show the result of our RG-calculation, and the 
line shows the result of the fit function defined in
Eq.~\eqref{eq:tcmpi_exp}.} 
\end{figure}

In Tab.~\ref{tab:tcmc_infvol} and 
Fig.~\ref{cap:Chiral-phase-transition}, 
we show the chiral phase transition 
temperature $T_{c}$ obtained in this way as a function of
the pion mass $m_{\pi}^{(0)}$.\footnote{We do not show lattice 
results for comparison in this figure since there is no 
data available for $m_{\pi}^{(0)}~\leq~300~\,~\mathrm{MeV}$ 
in Ref.~\cite{Karsch:2000kv}.} 
We find that the transition temperature
$T_{c}$ depends on the pion mass in the following way, 
\begin{equation}
T_{c}(m_{\pi}^{(0)}) = a_{0}+a_{1}m_{\pi}^{(0)} +
a_{2}(m_{\pi}^{(0)})^{2} + \mathcal{O}((m_{\pi}^{(0)})^{3}), 
\label{eq:tcmpi_exp}
\end{equation}
where the parameters can be determined from a fit to our numerical results as
\begin{equation}
a_{0} = 149.58\,\mathrm{{MeV}},\quad a_{1}=0.24258, \quad
a_{2}=0.00029\, \mathrm{{MeV}}^{-1}\,.
\end{equation}
$a_{0}$ is then the value for the chiral phase transition
temperature in the chiral limit as obtained from the fit. 
A similar relation was also found in lattice
simulations \cite{Karsch:2000kv, Bernard:2004je} with two or three
quark flavors. The corresponding relation is
\begin{equation}
\label{eq:tcmpi_karsch}
\frac{T_{c}(N_{f},\,
  m_{PS})}{\sqrt{\bar{\sigma}}}=\frac{T_{c}(N_{f},\,
  m_{PS}=0)}{\sqrt{\bar{\sigma}}} +
  l_{1}(N_{f})\frac{m_{PS}^{2/\beta\delta}}{\sqrt{\bar{\sigma}}}
  +\mathcal{O}(m_{PS}^{2})\,,
\end{equation}
where $m_{PS}$ denotes the mass of the pseudoscalar meson, and the string 
tension $\bar{\sigma}$ is used to set the scale in the lattice calculation. 
$\beta$ and $\delta$ are
the critical exponents of the $O(4)$-model in three dimensions. 
The coefficient $l_{1}(N_{f})$ depends slightly on the number of 
quark flavors \cite{Karsch:2000kv}.

\begin{table}
\begin{tabular}{|c||c|c|c|c|c|c|c|c|c|c|c|c|c|}
\hline 
$m_{\pi}\,\mathrm{[{MeV}]}$&
0&
30&
50&
75&
100&
125&
150&
175&
200&
225&
250&
275&
300\\
\hline
\hline 
$T_{c}\,[\mathrm{{MeV}]}$&
147.6&
157.5&
163.9&
170.5&
178.1&
184.2&
191.8&
200.2&
208.3&
218.1&
228.0&
238.5&
249.3\\
\hline 
$R_{c}(m_{\pi}^{(0)})$&
0&
0.067&
0.104&
0.155&
0.207&
0.250&
0.300&
0.356&
0.411&
0.478&
0.545&
0.616&
0.689\\
\hline
\end{tabular}
\caption{\label{tab:tcmc_infvol}Dependence of $T_{c}$ and
  $R_{c}(m_{\pi}^{(0)})=\frac{T_{c}(m_{\pi}^{(0)})-T_{c}(0)}{T_{c}(0)}$ 
on $m_{\pi}^{(0)}$.} 
\end{table}

The analysis in Eq.~\eqref{eq:tcmpi_karsch} assumes that the transition 
falls into the $O(4)$ universality class, where the ratio of the
critical exponents obeys $1/\beta\delta=0.55$.  
Then, the first order correction term is approximately linear, in
agreement with our result. 
On the lattice, however,
the coefficient of the approximately linear term is about one order of
magnitude smaller than the result of our calculation. For $N_f=3$, 
a lattice QCD calculation  \cite{Karsch:2000kv} gives $l_1
(N_f=3)\approx0.039$. While the exact value for $l_1 (N_f =2)$ is not given 
in \cite{Karsch:2000kv}, the authors point out that it is of 
the same order of magnitude as the value for $N_f =3$.

As we will see in the next section, it is not possible to
explain the smaller value of the (approximately) linear term found on
the lattice as 
a finite volume effect: Since a finite volume effect is more severe for
smaller pion masses and since it leads to a significantly reduced
transition temperature in our model, we expect that the slope of
$T_{c}(m_\pi ^{(0)})$ should actually increase in a finite volume, compared to the
infinite-volume result. We think that
the discrepancy may be a consequence of neglecting the 
gauge sector in the quark-meson model. 
In the chiral limit, the chiral phase transition temperature on
the lattice is about 
$30\,\mathrm{{MeV}}$ larger \cite{Karsch:2000kv}
than the value obtained in the quark-meson model.

Work on the quark-meson model within the Functional RG suggests 
that the transition temperature becomes even smaller if one 
includes wave function renormalizations \cite{Berges:1997eu}.
In spite of this, the slope of the function $T_{c}(m_{\pi} ^{(0)})$ is roughly
the same as in our study. This is an additional hint that neglecting the
gauge degrees of freedom could indeed be responsible for the
difference in the results from the quark-meson model compared 
to lattice calculations. 

A recent study in terms of the Functional RG (FRG), which incorporates
gluonic degrees of freedom and four-fermion interactions, shows reasonable
agreement with results from lattice studies of the chiral phase
transition temperature for two and three massless 
quark-flavors \cite{Braun:2005uj}.

Results for $T_{c}$ in the chiral limit
from various lattice and RG approaches are summarized in
Tab.~\ref{tab:tctable}. As can be seen from the table,
there is some uncertainty in the value of the chiral phase 
transition temperature in lattice calculations, which is mainly due to 
different implementations of the fermions. 

\begin{table}
\begin{tabular}{|l|l|c|}
\hline 
Reference & Method & $T_{c}$ [MeV] \\
\hline
\hline 
this work  & Proper-time RG & $ 148$ \\ 
\hline
Berges 
(1997) \hfill \cite{Berges:1997eu} & Functional RG, quark-meson model &
$100.7$ \hfill \\
\hline
Schaefer (1999) \hfill \cite{Schaefer:1999em} & Proper-time RG, quark-meson model & $149$ \\ 
\hline
Braun 
(2003) \hfill \cite{Braun:2003ii} & Proper-time RG, quark-meson model & $154$ \\
\hline
Schaefer (2004) \hfill \cite{Schaefer:2004en} & Proper-time RG, quark-meson model & $142$
\\ 
\hline 
Braun (2005) \hfill \cite{Braun:2005uj} & Functional RG, QCD (see caption) & $186$  
\\
\hline
Gottlieb 
(1996) \hfill \cite{Gottlieb:1996ae} & lattice $16^3
\times 8$ (staggered) & $128 \pm 9$ \\
\hline
Karsch 
(2000) \hfill \cite{Karsch:2000kv} & lattice $16^3 \times
4$ (improved staggered) & $173 \pm 8$ \\ 
\hline  
CP-PACS (2000) \hfill \cite{AliKhan:2000iz} & lattice $16^3 \times 4$
(improved Wilson) & $171 \pm 4$ \\
\hline 
Bornyakov 
(2005) \hfill \cite{Bornyakov:2005dt} & lattice
$16^3 \times 8$ (improved Wilson) & $173 \pm 3$ \\
\hline
\end{tabular}

\caption{\label{tab:tctable} Chiral phase transition temperature 
in the chiral limit ($m_\pi \to 0$) from different RG approaches 
for the quark-meson model and for QCD, and from lattice simulations. 
We have restricted our choice of lattice references to the case 
of $N_f=2$ flavors that we have treated here. More recent lattice results have
been obtained for 
$N_f=2+1$ flavors, see e.g. \cite{Karsch:2001nf, Fodor:2004nz, Bernard:2004je}.
The difference in the 
RG results arises from a weak dependence of $T_{c}$
on the initial values at the UV scale and on the choice of the cutoff-function
Eq.~\eqref{eq:cutoff-fct}. In Ref.~\cite{Braun:2005uj}, $T_c$  
is calculated from a study which incorporates the running QCD coupling 
and all possible four-fermion interactions in the chiral symmetric regime.} 
\end{table}

\section{Chiral Phase Transition Temperature in Finite Spatial Volumes}
\label{sec:cpt_fv}

\begin{figure}
\includegraphics[clip, scale=1.2]{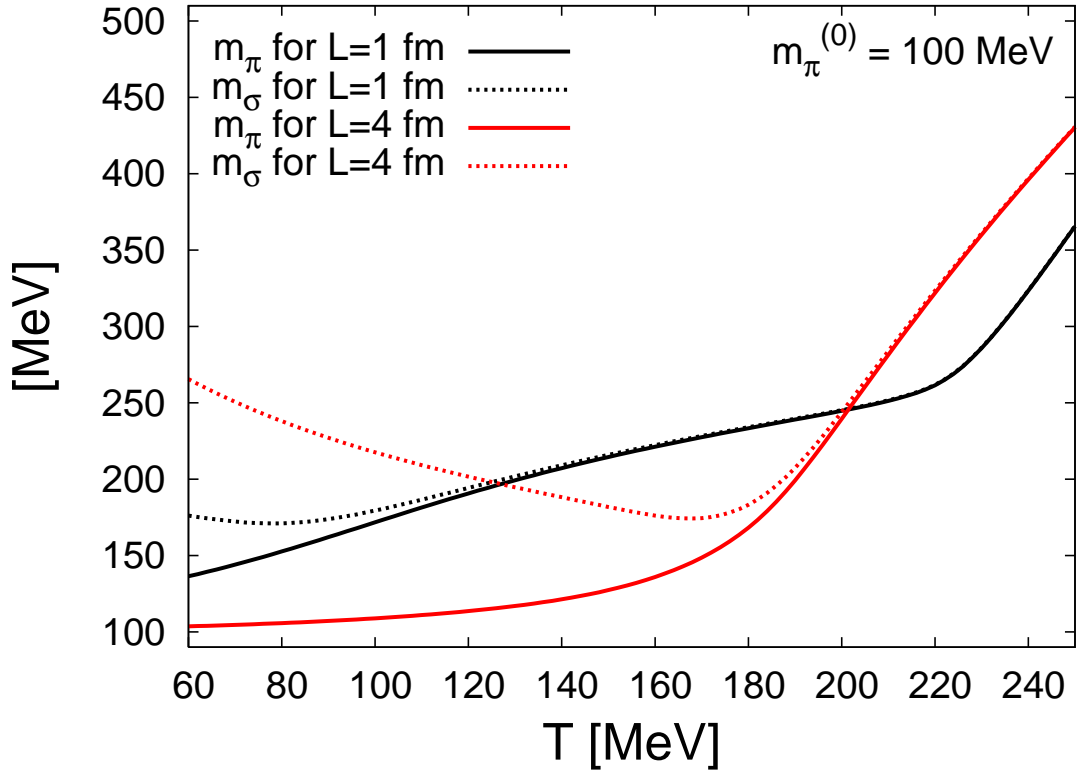}

\caption{\label{cap:sigmamass}Sigma- and Pion-mass as functions of the
  temperature $T$, with $m_{\pi}^{(0)}=100\;\mathrm{MeV}$ and periodic
quark boundary conditions, for $L=1\;\mathrm{{fm}}$
and $L=4\;\mathrm{{fm}}$.
The solid black and the red/gray lines show the pion mass
for $L=1\mathrm{{fm}}$ and $L=4\,\mathrm{{fm}}$, respectively, whereas 
the dotted black and red/gray lines show the sigma mass 
for $L=1\; \mathrm{{fm}}$ and for $L=4\;\mathrm{{fm}}$.}
\end{figure}

We now turn to the investigation of the chiral phase transition temperature
in finite spatial volumes. As in section~\ref{sec:cpt_iv}, 
we define the phase transition temperature $T_{c}$ 
via the minimum of the $\sigma$-mass. 
Putting the system in a finite volume introduces an additional
scale. Let us first discuss the influence of this additional length 
scale $L$ on the $\sigma$-mass and $\pi$-mass. In Fig.~\ref{cap:sigmamass},
we show the $\sigma$-mass and $\pi$-mass for
$m_{\pi}^{(0)}=100\,\mathrm{{MeV}}$ and with periodic boundary
conditions for the quarks, 
as a function of the temperature, for both a small volume
$L=1\,\mathrm{{fm}}$ and a large volume $L=4\,\mathrm{{fm}}$. The 
minimum of the $\sigma$-mass is clearly visible in the plot. Above the 
transition temperature $T_{c}$, where chiral symmetry
is restored, the $\sigma$- and $\pi$-mass are degenerate, independent of 
the size of the volume.
In order to gain a better
understanding of the meson masses and their dependence on the scales
$L$ and $T$ in this temperature regime, we peform a perturbative one-loop
calculation. Since the chiral phase is a non-perturbative 
phenomena, such a one-loop calculation is not
  trustworthy in the vicinity of the critical temperature: The extraction of  
the critical temperature fails, leading to an unphysical complex temperature
\cite{Dolan:1973qd}. Here, higher-loop terms contribute significantly, as
pointed out by earlier RG flow studies, eg. Ref.  \cite{Berges:1997eu,
  Schaefer:1999em, Braun:2003ii},  and a study in terms 
 of many-body resummation techniques \cite{Dumitru:2003cf}.
 Moreover, we neglect the quark contributions in this calculation, since 
they are suppressed by the appearance of a thermal Matsubara mass.  
In contrast, the bosonic fields have vanishing Matsubara mass and therefore 
their contributions dominate at high temperature.
Thus our starting point for the calculation of the mass correction is
the scalar $O(N=N_f ^2)$-model with the Lagrangian
\begin{equation}
\label{eq:ON_lag}\mathcal{L} = \frac{1}{2}(\partial_{\mu}\phi)^{2} +
\frac{1}{2}m^{2}\phi^{2} + \frac{\lambda}{4}\phi ^{4}\,, 
\end{equation}
where we have introduced the $O(N)$-vector $\phi=(\phi _1,...,\phi
_N)$, and the parameters are $m=m_{UV}$ and $\lambda=\lambda_{UV}$.

The mass correction $\delta m^{2}(T,L)$, which is due to finite volume and
finite temperature effects, can be decomposed into a sum of two
contributions, 
$\delta m_{1}^{2}(T,L\!\rightarrow\!\infty)$ and $\delta
m_{2}^{2}(T,L)$. We refer to Appendix~\ref{sec:oneloop} 
for details of the calculation. 

First, in the regime defined by $0<\frac{1}{T}\ll L$, the contribution $\delta
m_{1}^{2}(T,L\!\rightarrow\!\infty)$ 
dominates. One can estimate $\delta m_{1}^{2}(T,L\!\rightarrow\!\infty)$ 
for large temperatures and volumes as
\begin{equation}
\delta m_{1}^{2}(T ,L\!\rightarrow\!\infty) \approx(N+2)\frac{\lambda}{12}T^{2}
\, \; \mbox{for} \;\; T \to \infty. \label{eq:dm2_T}
\end{equation}
In this case, the meson masses depend linearly on the
temperature, in agreement with the result from Ref.~\cite{Dolan:1973qd}. 
 
Second, if $T$ and $L$ are of the same
order of magnitude, the mass correction is in essence given by
\begin{equation}
\delta m_{2}^{2}(T,L) \approx \frac{(N+2)\lambda
  }{(2\pi)^{\frac{3}{2}}} \frac{T}{L} \sum_{n=-\infty}^{\infty} \sum_{\{
  l_{i}\}} \!'
  \Big(\frac{(mL)^{2}+4\pi^{2}n^{2}(TL)^{2}}{\vec{{l}}^{2}}\Big)^{\frac{1}{4}}
  K_{\frac{1}{2}} \Big(
  \sqrt{\vec{{l}}^{2}((mL)^{2}+4\pi^{2}n^{2}(TL)^{2})} \Big), 
  \label{eq:dm2_TL}
\end{equation}
where $K_{n}$ denotes the modified Bessel-functions with index
$n=\frac{1}{2}$. The vector $\vec{l}$ is defined as $\vec{{l}}=\{
l_{1},l_{2},l_{3}\}$ and  
the prime indicates that the term with $\vec{{l}}=0$ is excluded from
the summation. 
Note that $\delta m_{2}^{2}(T,L)$ has a complicated dependence
on $T$ and $L$, but we observe that it scales with $\frac{T}{L}$,
rather than with $T^2$.  
This explains the difference between the slopes of the meson masses in
the regime defined by $\frac{1}{T_{c}}>\frac{1}{T}\gtrsim L$, and in the
regime defined by $0<\frac{1}{T}\ll L$, which can be seen in
Fig.~\ref{cap:sigmamass}. 

In contrast, in the limit $TL\gg 1$ one obtains 
\begin{equation}
\delta m_{2}^{2}(T,L) \approx \frac{3(N+2)\lambda}{\sqrt{8}\pi}
\frac{T}{L} \sum_{n=-\infty}^{\infty} \sum_{\{
  l_{i}\}}{}^\prime \frac{1}{\sqrt{\vec{l}^2}} \exp \Big(
-\sqrt{\vec{{l}}^{2}((mL)^{2}+4 \pi^{2} n^{2} (TL)^{2})}\Big).
\label{eq:dm2_TL2}
\end{equation}
The contributions from the non-vanishing thermal Matsubara-modes
to $\delta m_{2}^{2}(T,L)$ drop exponentially, and $\delta m_{2}^{2}(T,L)$
becomes a linear function in the temperature $T$, due to the zeroth
thermal Matsubara-mode $n=0$. Therefore, for $TL \gg 1$, $\delta
m_{2}^{2}(T,L)$ is a sub-leading correction to the meson masses,
compared to the contribution $\delta m_{1}^{2}(T,L)$. This describes
the results for periodic quark boundary conditions well.

\begin{figure}
\includegraphics[scale=0.4]{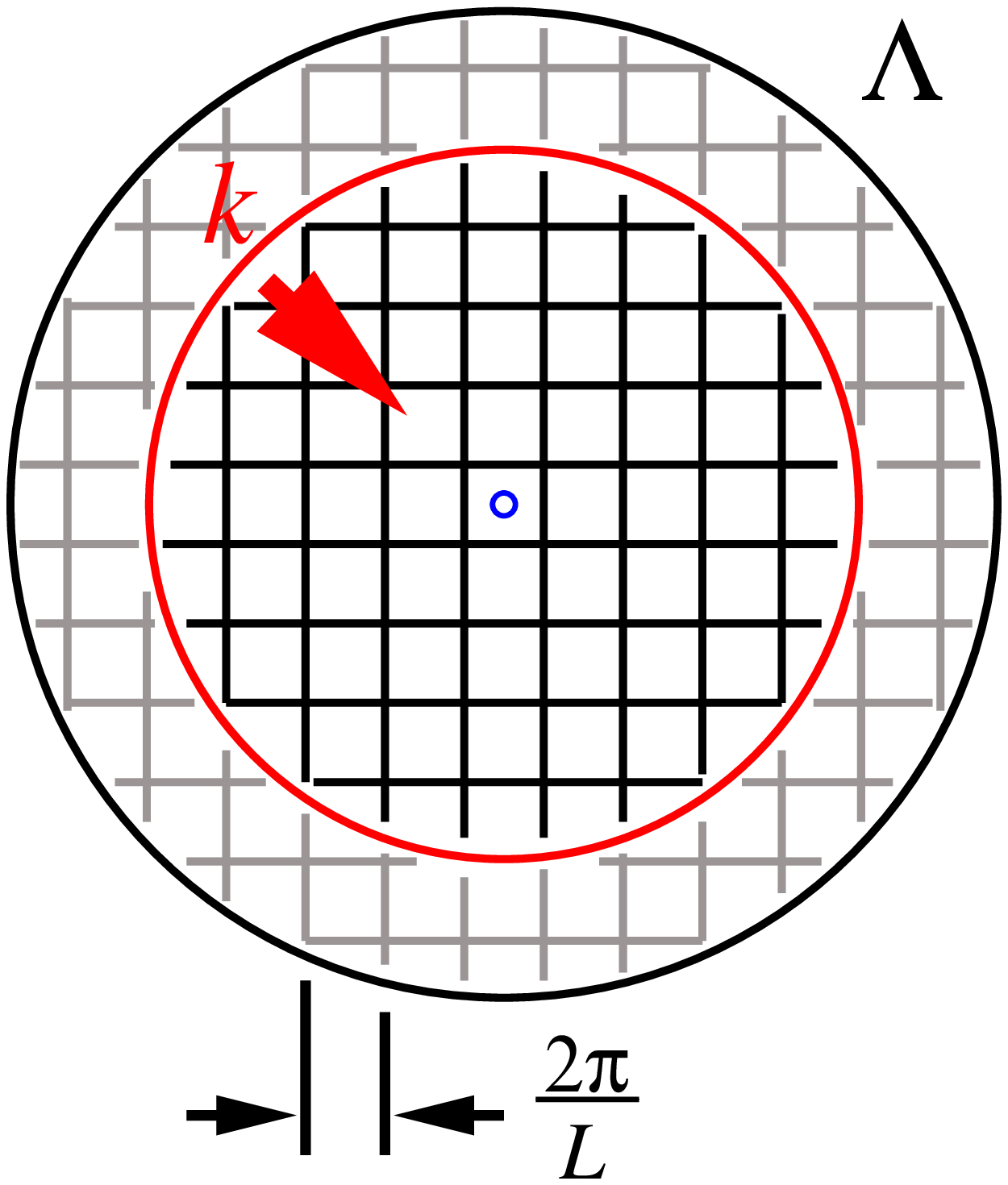}
\hspace{2cm}\includegraphics[scale=0.4]{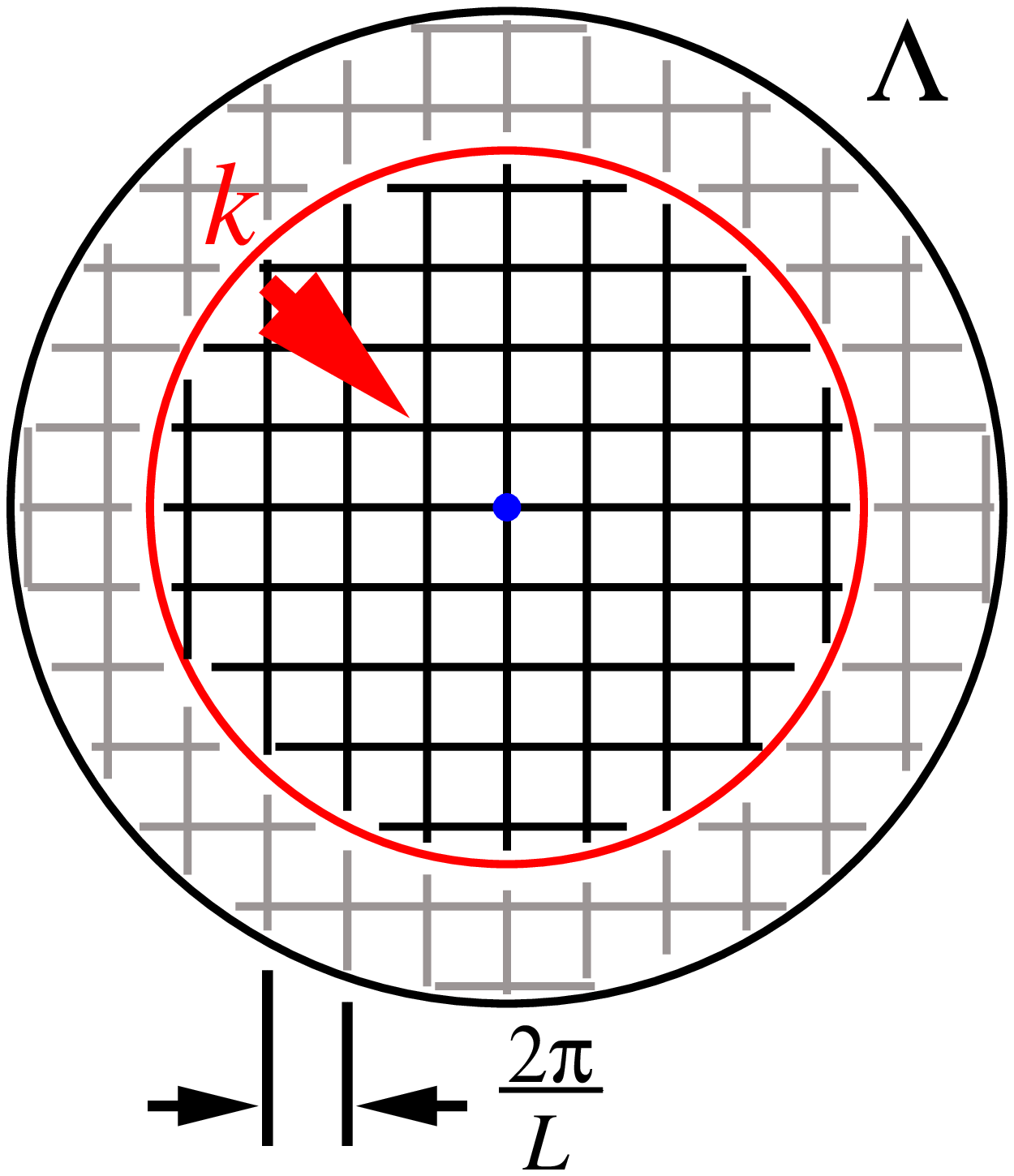}
\caption{\label{cap:schematical} Schematic representation of the relation
between the momentum summation and the UV- and IR-cutoff in momentum
space, for anti-periodic boundary conditions (left panel) and for periodic
boundary conditions (right panel). The UV-cutoff is denoted by $\Lambda$,
while $k$ denotes the variable IR-cutoff of the RG scheme. The arrow
indicates the direction of 
the RG flow from $k = \Lambda_{UV}$ to $k = 0$.}
\end{figure}

Similar behavior is found for
anti-periodic boundary conditions of the quark fields. However, there is one 
essential difference between periodic and anti-periodic
boundary conditions: In the case of
anti-periodic boundary conditions, the quark fields have a finite minimal
infrared momentum
\begin{equation}
p_{ap}^{\mathrm{min}}=\frac{\pi}{L}\,, 
\end{equation}
which is illustrated in Fig. \ref{cap:schematical}. In the
quark-propagator, the minimal value $p_{ap}^{\mathrm{min}}$ acts as an
additional mass term which increases for decreasing volume sizes.
The quark fields decouple from the RG flow as soon as the IR-cutoff
scale $k$ in Eq.~\eqref{eq:FV_fe} drops below
$p_{ap}^{\mathrm{min}}$. Consequently, the mesons 
are the only dynamical degrees of freedom in the theory for $k\leq
p_{ap}^{\mathrm{min}}$: The system becomes equivalent to an
$O(4)$-model, which remains in the symmetric phase for $k\rightarrow 0$.

\begin{figure}
\includegraphics[clip,scale=0.81]{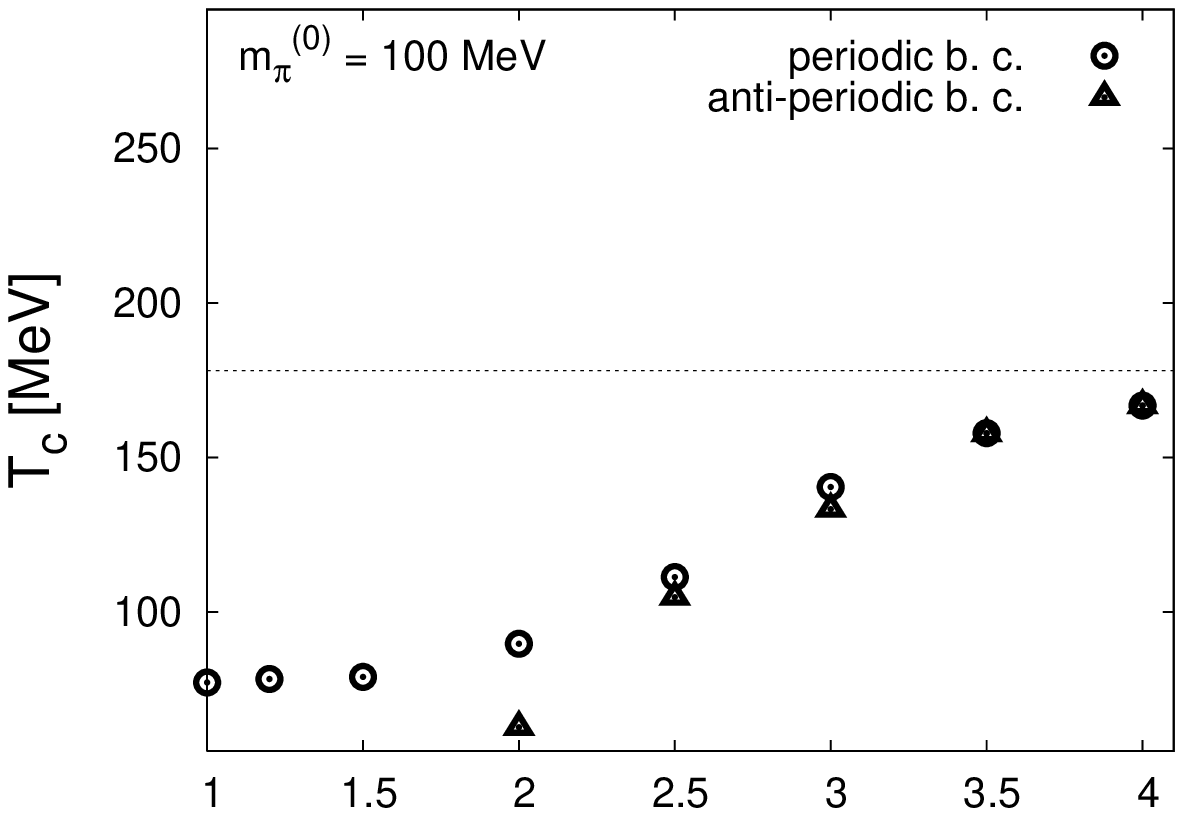}\\
\includegraphics[clip,scale=0.81]{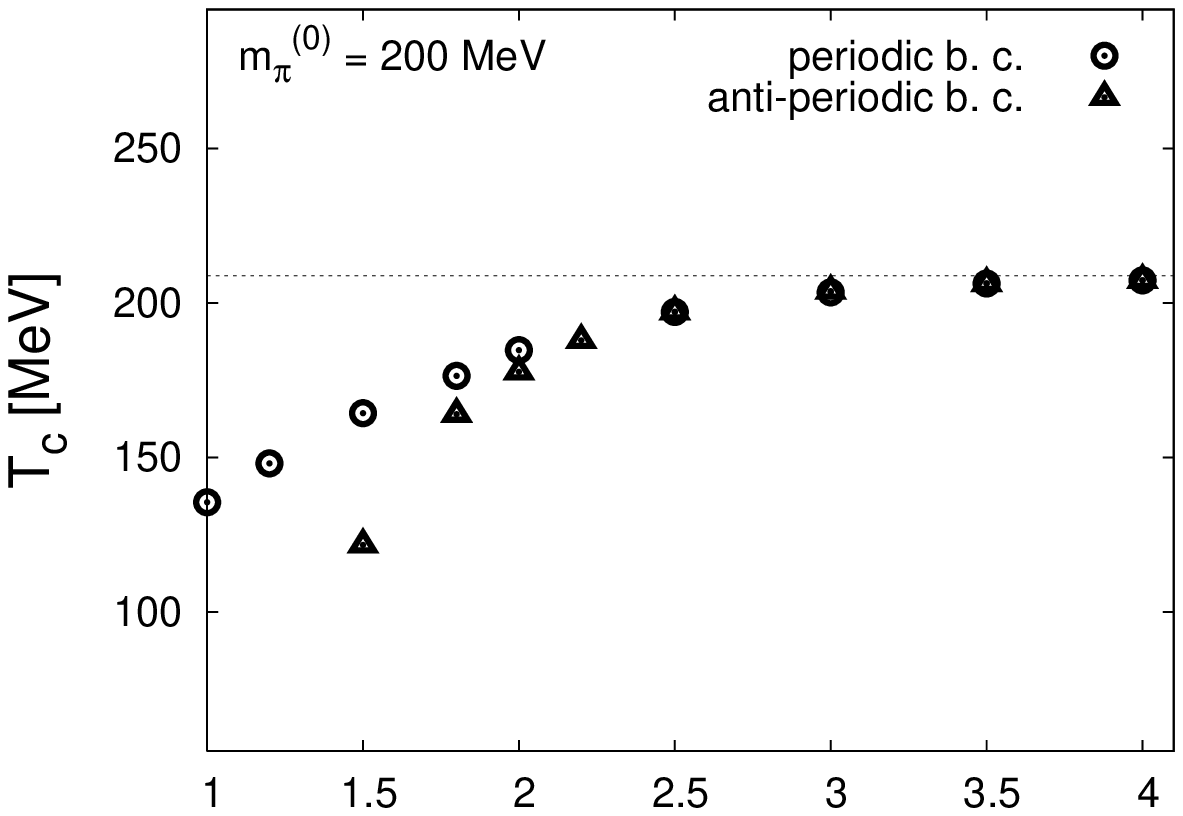}\\
\includegraphics[clip,scale=0.81]{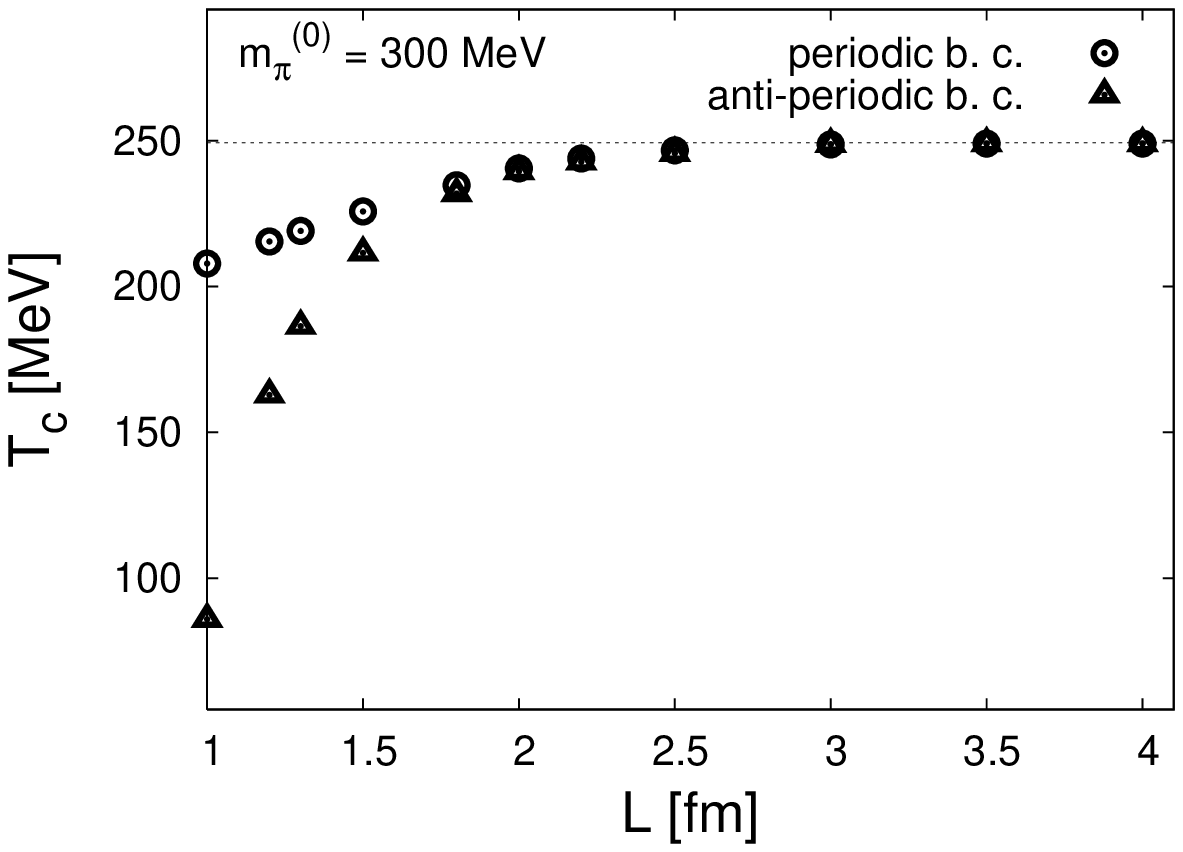}
\caption{\label{cap:TcL} The chiral phase transition temperature $T_{c}$ as a
function of the volume size $L$, for different pion masses $m_{\pi}^{(0)}$, 
and for both choices of spatial quark boundary conditions. 
Results for periodic (anti-periodic) boundary conditions
are denoted by circles (triangles), values for the pion mass at
$T=0$ in infinite volume are given in the panels.}
\end{figure}

In general, the momentum scale $\frac{2\pi}{L}$ should be much smaller 
than the UV cut off $\Lambda _{UV}$,
\begin{equation}
p_{p} ^{L}=\frac{2\pi}{L}\ll\Lambda_{UV}\,.
\end{equation}
If $p_{p}^{L}$ becomes comparable to $\Lambda _{UV}$, there are 
no dynamical degrees of freedom left in our model.

We now present our main results for the volume dependence
of the chiral phase transition temperature. 
Fig.~\ref{cap:TcL} contains plots of the transition temperature
$T_{c}$ as a function of the volume size, for different values of the pion mass
at zero temperature, $m_{\pi}^{(0)}$, and for different choices for the
quark boundary conditions. 
For small, realistic pion masses, $m_{\pi}^{(0)}=100\,\mathrm{MeV}$,
already for  
$L=4\,\mathrm{{fm}}$ the
results in the upper panel of Fig.~\ref{cap:TcL} show a deviation of
$T_{c}$ from 
its infinite volume value of about 6\%, independent of the choice of
boundary conditions.  
For small volume sizes $L < \frac{1}{m_{\pi}^{(0)}}$, $T_{c}$ is 
strongly affected by the choice of the boundary conditions
for the fermions. For anti-periodic boundary conditions, we observe
that $T_{c}$ decreases strongly for small volume sizes.
As already discussed above, this is because of the additional 
infrared cutoff $p_{ap}^{\mathrm{min}}$ for the momenta of the quark fields.
It is due to this additional IR-cutoff for the quarks
that the system remains in the symmetric phase for small 
volume sizes \cite{Braun:2004yk, Braun:2005gy}.

For periodic quark boundary conditions, we observe
a weaker volume dependence of $T_{c}$, since the condensation of the
quarks is not prevented by the additional IR-cutoff $p_{ap}^{\mathrm{min}}$.
For $m_{\pi}^{(0)}=100\,\mathrm{{MeV}}$ and $L \lesssim
1.5\,\mathrm{{fm}}$, we  observe that $T_{c}$ is almost independent of
$L$, which may be due to the fact that $p_{p}^{L}$
approaches $\Lambda _{UV}$. 

For large pion masses, $m_{\pi}^{(0)}\gtrsim 300\,\mathrm{MeV}$, and
for $L\geq 2\,\mathrm{{fm}}$, $T_{c}$ depends only weakly on the box size.
The deviation from its infinite volume value is less than $1 \%$
already for $L\approx 2.5\,\mathrm{{fm}}$.  
We observe only a weak dependence
on the choice of the fermionic boundary conditions, as well. The reason is that
the length scale set by the pion mass,
$L_{\pi}\sim\frac{1}{m_{\pi}^{(0)}}$, is much smaller than 
the box size $L$. Therefore the volume dependence of $T_c$ is governed by the 
pion mass scale $m_{\pi}^{(0)}$, rather than the scale set by the spatial box
size: pion fluctuations are more strongly suppressed by their large
mass than by the long-wavelength cutoff from the finite volume. This
observation implies that lattice results    
for $T_c$ are not affected by the finite volume to any considerable
degree, provided the pion mass is large, $m_{\pi}^{(0)}\gtrsim
300\,\mathrm{MeV}$. 

\begin{figure}
\includegraphics[clip,scale=0.9]{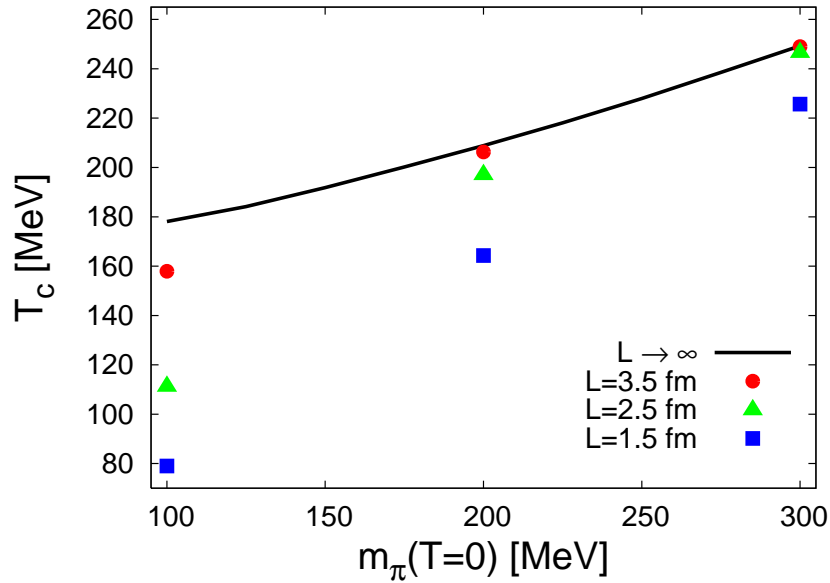}
\caption{\label{cap:tc_mpi_vol}Chiral phase transition temperature
$T_{c}$ as a function of $m_{\pi}(T=0)$ for different box sizes. 
We show the results for $L=1.5,\,2.5,\,3.5\;\;\mathrm{fm}$ 
from bottom to top. For comparison, we also show the 
results for infinite volume (solid line) from 
Fig.~\ref{cap:Chiral-phase-transition}.}
\end{figure}

Finally, we stress that finite volumes make the coefficient $a_{1}$ in
Eq.~\eqref{eq:tcmpi_exp} bigger for smaller volumes. 
This can be seen from Tab.~\ref{tab:Tc} and 
Fig.~\ref{cap:tc_mpi_vol}, where the slope of $T_c(m_\pi ^{(0)}, L)$ as a function of $m_\pi^{(0)}$ is even
larger at smaller values of $L$.

\begin{table}
\begin{tabular}{|c||c|c|c|}
\hline 
$L \;\mathrm{[fm]}$ &
$m_\pi^{(0)}=100\,\mathrm{{MeV}}$&
$m_\pi^{(0)}=200\,\mathrm{{MeV}}$&
$m_\pi^{(0)}=300\,\mathrm{{MeV}}$\\
\hline
\hline 
$1.5$&
79.0 MeV&
164.3 MeV&
225.7 MeV\\
\hline 
$2.5$&
111.3 MeV&
197.1 MeV&
246.6 MeV\\
\hline 
$3.5$&
157.9 MeV&
206.3 MeV&
249 MeV\\
\hline
\hline 
$\infty$&
178.1 MeV&
208.3 MeV&
249.3 MeV\\
\hline
\end{tabular}
\caption{\label{tab:Tc} Chiral phase transition temperatures
  $T_c(m_\pi^{(0)}, L)$ as function
of the pion mass $m_\pi^{(0)}$
and the box size $L$ for periodic boundary conditions for the quark-fields.
For comparison, the corresponding values in infinite spatial volume 
are also given.}
\end{table}

\section{Conclusions}
\label{sec:conclusions}

In this article, we have presented results for the volume dependence of the
phase transition temperature for the chiral symmetry restoration
transition. Our investigation is based on the quark-meson model and
uses Renormalization Group methods. In  
this way, we obtain non-perturbative results for the transition
temperature for various values of the pion mass. 

In general, no phase transition can occur in a system of finite
volume. The evaluation of lattice results therefore uses a scaling analysis in
quark mass and temperature \cite{Karsch:1994hm, Aoki:1998wg, Bernard:1999xx,
  D'Elia:2005bv}, and recently finite-size scaling
\cite{D'Elia:2005bv} as an analytical  tool. We expect that a 
Renormalization Group analysis of the critical behavior can complement
these approaches.

In the present paper, we have focused on the chiral phase transition
temperature, which is not universal and also model-dependent.
However, the relative shift of the temperature from
infinite to finite volume should depend mainly on the pion mass and
the pion decay constant, which are independent of the model and represent an
external input to our calculations. 

We find that finite volume effects for the
transition temperature remain small for large pion masses
$m_{\pi}\gtrsim 300\,\mathrm{{MeV}}$, as long as the volume is of the
order $L \ge 2\,\mathrm{{fm}}$ in the spatial directions. 
The scale for the appearance of sizable finite-volume effects is given 
by the pion mass $m_{\pi}$, and 
the effects remains small as long as $L \gg \frac{1}{m_{\pi}}$. 

On the other hand, finite size effects are
sizable already at a lattice extent of $L \simeq 2 \, \mathrm{fm}$
for realistic pion masses of the order of $100 \, \mathrm{MeV}$. We expect 
therefore that finite volume effects will become more 
relevant in future simulations with realistic pion masses.
The strategy for lattice calculations should then be to
simulate in volumes where the value $m_\pi L$ is large enough to keep the 
finite volume effects down to an acceptable size, and to extrapolate
to smaller pion masses and the chiral limit. 

We note that the choice of periodic boundary conditions in spatial
directions for the quark fields, which is commonly employed in lattice
simulations, leads to a much smaller finite volume effect on the
transition temperature. This conclusion agrees with our results for 
the volume dependence of the pion mass and pion decay constant
\cite{Braun:2005gy}. 

The dependence of the transition temperature on the quark mass in the
quark-meson model is much stronger than that observed in lattice
simulations. This must remain a puzzle in the present approach and
calls for a more extensive consideration of the gluon dynamics. Work
in this direction has already been started \cite{Braun:2005uj}. 
In concert with other non-perturbative
methods, the Renormalization Group approach should prove to be valuable as a
complement to QCD lattice calculations.

\acknowledgments

JB would like to thank the GSI for financial support. This work is
supported in part by the Helmholtz association under grant
no. VH-VI-041, and in part by the EU Integrated Infrastructure
Initiative Hadron Physics (I3HP) under contract RII3-CT-2004-506078.  
AHR acknowledges the financial support of the Alexander von Humboldt 
foundation.
\appendix

\section{\label{sec:Starting-values}Initial Values at the UV-Scale}

\begin{table}
\begin{tabular}{|c|c||c|c|}
\hline 
$m_{UV}\,\mathrm{[{MeV}]}$&
$gm_{c}\,\mathrm{[{MeV}]}$&
$m_{\pi}(k\rightarrow0)\,\mathrm{[{MeV}]}$&
$f_{\pi}(k\rightarrow0)\,[\mathrm{{MeV}}]$\tabularnewline
\hline
\hline 
794&
\phantom{0}0.036&
\phantom{0}30&
\phantom{0}87\tabularnewline
\hline 
792&
\phantom{0}0.39\phantom{0}&
\phantom{0}50&
\phantom{0}87\tabularnewline
\hline 
788&
\phantom{0}1.08\phantom{0}&
\phantom{0}75&
\phantom{0}88\tabularnewline
\hline 
779&
\phantom{0}2.10\phantom{0}&
100&
\phantom{0}90\tabularnewline
\hline 
775&
\phantom{0}3.43\phantom{0}&
125&
\phantom{0}91\tabularnewline
\hline 
767&
\phantom{0}5.15\phantom{0}&
150&
\phantom{0}93\tabularnewline
\hline
757&
\phantom{0}7.275&
175&
\phantom{0}95\tabularnewline
\hline
748&
\phantom{0}9.85\phantom{0}&
200&
\phantom{0}97\tabularnewline
\hline
737&
12.93\phantom{0}&
225&
\phantom{0}99\tabularnewline
\hline
725&
17.00\phantom{0}&
250&
101\tabularnewline
\hline
711&
20.80\phantom{0}&
275&
103\tabularnewline
\hline
698&
25.70\phantom{0}&
300&
105\tabularnewline
\hline
\end{tabular}
\caption{\label{cap:sv}Boundary conditions $a_{ij}(\Lambda_{UV})$ for
  the differential equations for the coefficients $a_{ij}(k)$. 
The corresponding values for $m_{\pi}$ and $f_{\pi}$ which result from
the RG flow at zero temperature and in infinite 
volume are shown in the last two columns. The coefficients $a_{01}$
  and $a_{02}$, which are initially
non-zero, can be expressed in terms of the two parameters $m_{UV}$ and
$\lambda_{UV}$, for the relation see text. As an initial condition for
the flow equation 
for $\lambda(k)$, we have chosen $\lambda(\Lambda_{UV})=\lambda_{UV}= 60$ 
for all parameter sets. The initial conditions for the differential
equations for the remaining coefficients $a_{ij}(k)$
of the potential Eq.~\eqref{eq:pot_ansatz} are
$a_{ij}(\Lambda_{UV})=0$. In addition, we use $g=3.258$ for the
Yukawa-coupling. In our notation, $g m_c$ is the current quark mass.} 
\end{table}

In this section, we give an overview of the initial values for the
flow equations of the coefficients $a_{ij}(k)$. As already discussed
in Sec.~\ref{sec:flow-equations}, we fix the values of the
coefficients $a_{ij}(\Lambda_{UV})$
at the scale $\Lambda_{UV}$ in the infinite four-dimensional Euclidean
volume in such a way that the values of $m_{\pi}$ and $f_{\pi}$
given by chiral perturbation \cite{Colangelo:2003hf} are
reproduced. However, some freedom remains in the choice of the
starting values for the 
coefficients. But as discussed in \cite{Braun:2005gy}, different
sets of parameters give the same dependence on the size of the four
dimensional volume, provided that they lead to the same values of
the pion decay constant and pion mass in infinite volume. The values
which we have used for our calculation can be determined from
Tab.~\ref{cap:sv}. In this table, we give the value for the parameter
$m_{UV}$. Together 
with the parameter $\lambda_{UV}=\lambda(\Lambda_{UV})$ and the
evolving minimum of the potential $\sigma_0(k)$, the two coefficients
that are initially non-zero can be expressed as 
\be
a_{01}(\Lambda_{UV}) &=&
\frac{1}{2}(m_{UV}^2+\lambda_{UV}\sigma_{0}^2(\Lambda_{UV})) \\
a_{02}(\Lambda_{UV}) &=& 
\frac{1}{4} \lambda_{UV}\,.
\ee
All other coefficients are initially set to zero.
In principle, one should choose $\sigma_{0}(\Lambda_{UV})=0$ as
initial condition, but in order to
avoid numerical problems at the UV-scale, we choose a small but finite
value for $\sigma_{0}$ at the UV-scale,
e.g. $\sigma_{0}(\Lambda_{UV})=0.1\,\mathrm{{MeV}}$. 
The parameters $m_{UV}$ and $\lambda_{UV}$ can be
interpreted as the usual meson masses and four-point couplings at the
UV scale.
The meson masses are related to the coefficients $a_{ij}$ and the
minimum $\sigma_{0}$ of the potential 
by \cite{Braun:2004yk}
\begin{eqnarray}
m_{\pi}^{2} & = & 2a_{01}\\
m_{\sigma}^{2} & = &
m_{\pi}^{2}+2a_{20}+4a_{11}\sigma_{0}+8a_{02}\sigma_{0}^{2}\end{eqnarray} 

\section{\label{sec:oneloop}One-Loop Calculation of the Meson Masses}

In order to gain a better understanding of the slope of the meson masses
as a function of temperature in the symmetric phase, we compared the
RG results to a
one-loop calculation for the mass of a scalar $O(N)$-model, as defined
by the Lagrangian in Eq.~\eqref{eq:ON_lag}. More details of the
calculation are given here.

The effective potential
in one-loop approximation in a $d$-dimensional Euclidean space reads
\be
U^{\mathrm{{1L}}}(\phi) &=& \frac{1}{2} \Tr\log
(-\partial^{2}+M^{2}(\phi))\nn\\
&=& -\frac{1}{2} \frac{T}{L^{d-1}} \sum_{j=1}^{N} \sum_{\{ n_{i}\}}
\int_{\frac{1}{\Lambda^{2}}}^{\infty} \frac{ds}{s} 
 \exp\Big( - s(4\pi^{2}T^{2}n_{0}^{2} + \frac{4 \pi^{2}}{L^{2}}
 \sum_{i=1}^{d-1} n_{i}^{2}+M_{j}^{2}(\phi)) \Big)\,,\label{eq:U1L}
\ee
where we have used Schwinger's proper time representation of the
logarithm, supplemented by a UV-cutoff $\Lambda$.
In addition, we have introduced the abbreviations 
$M_{1}^{2}(\phi)=m^{2}+3\lambda\phi^{2}$
and $M_{j}^{2}(\phi)=m^{2}+\lambda\phi^{2}$ for $j=2,...,N$.
The sum runs over all permutations of $\{ n_{1},...,n_{d}\}$. Here
$L$ denotes the length of the $d$-dimensional box in space directions,
and $\frac{1}{T}$ denotes the length of the box in the Euclidean
time direction. In order to calculate the mass of the scalar field,
we use the Jacobi-Elliptic-Theta function $\vartheta_{p}(s)$, defined
in Eq.~\eqref{eq:theta_p}. Both representations of this function
given in Eq.~\eqref{eq:theta_p} deserve some comments: The first
representation in Eq.~\eqref{eq:theta_p} is essentially the usual
Matsubara summation, where a truncation of the Matsubara sum can
be used to perform a high-temperature or small-volume expansion of
the one-loop effective potential. In order to get the second representation
in Eq.~\eqref{eq:theta_p}, we have applied Poisson's formula to the
first representation. This second representation allows to separate
the UV divergence of the effective potential: the divergent
infinite-volume contribution, which is given by the
first term of the left-hand side of Eq.~\eqref{eq:theta_p}, can be
separated from the volume and temperature dependent parts.
In addition, a truncation of the sum in the second representation can
be used to perform a low-temperature or large-volume expansion of the
one-loop effective potential.

With the second representation of the Jacobi-Elliptic-Theta function
from Eq.~\eqref{eq:theta_p}, we can divide the one-loop effective
potential in three contributions,
\begin{equation}
U^{\mathrm{{1L}}} =
U_{\infty}^{\mathrm{{1L}}}(T\!\rightarrow\!0,L\!\rightarrow\!\infty) +
U_{1}^{\mathrm{{1L}}}(T,L\!\rightarrow\!\infty) + U_{2}^{\mathrm{{1L}}}(T,L)\,.
\end{equation}
We do not consider the contribution $U_{\infty}^{\mathrm{{1L}}}$
in the following, since we are only interested in the finite-temperature
and finite-volume corrections to the mass. We
start with the calculation of the mass correction due to the
volume-independent contribution
$U_{1}^{\mathrm{{1L}}}(T) = U_{1}^{\mathrm{{1L}}}(T,L\!\rightarrow\!\infty)$
which is given by
\be
U_{1}^{\mathrm{{1L}}}(T) = -\frac{1}{(4\pi)^{\frac{d}{2}}}
\sum_{j=1}^{N} \sum_{q=1}^{\infty}\int_{0}^{\infty}
\frac{ds}{s^{1+\frac{d}{2}}} 
\exp \Big(-\frac{q^{2}}{4 s T^{2}}- s M_{j}^{2}(\phi)\Big) \,.
\ee
We stress that the regulator $\Lambda$ of the Schwinger proper-time
integral can be removed here, $\Lambda\rightarrow\infty$. In
order to calculate the mass correction in the symmetric phase, we
have to take the second derivative of $U_{1}^{\mathrm{{1L}}}(T)$
with respect to the field $\phi$ and evaluate the resulting expression
at $\phi=0$:
\begin{equation}
\delta m_{1}^{2}(T) =
\left.\frac{\partial^{2}}{\partial\phi^{2}} U_{1}^{\mathrm{{1L}}}(T)
\right|_{\phi=0} 
=\frac{2(N+2)\lambda}{(2\pi)^{\frac{d}{2}}} (mT)^{\frac{d-2}{2}}
\sum_{q=1}^{\infty} \frac{1}{q^{\frac{d-2}{2}}}\, 
K_{\frac{d-2}{2}}\big(\frac{qm}{T}\big)\,.\label{eq:dm1_2_bessel}
\end{equation}
Now we make use of the integral representation of the modified Bessel
Functions, 
\begin{equation}
2\left(\frac{x}{2}\right)^{\nu}K_{\nu}(x)
=\int_{_{0}}^{\infty}dss^{\nu-1}\E^{-s-\frac{\pi^{2}}{4s}}\,.\label{eq:bessel}
\end{equation}
For $T\rightarrow\infty$ and $d=4$, we use 
\begin{equation}
K_{1}(x)\approx\frac{1}{x}\quad(x \ll 1)\,,
\end{equation}
and obtain the simple expression
\begin{equation}
\delta m_{1}^{2}(T)\approx\frac{(N+2)\lambda}{12}T^{2}
\end{equation}
for the correction $\delta m_{1}^{2}(T)$, which agrees with the result
found in \cite{Dolan:1973qd}.

Now we turn to the calculation of the mass correction due to the contribution
$U_{2}^{\mathrm{{1L}}}(T,L)$. Since we are interested in small spatial
volumes and high temperatures, we use the Poisson-representation for
the spatial contributions, but we still use the 
usual Matsubara sum for the thermal contribution.
The contribution $U_{2}^{\mathrm{{1L}}}(T,L)$ to the potential is
\begin{equation}
U_{2}^{\mathrm{{1L}}}(T,L) = -\frac{T}{2(4\pi)^{\frac{d-1}{2}}}
\sum_{j=1}^{N} \sum_{n=-\infty}^{\infty} \sum_{\{ l_{i}\}} \!'
\int_{0}^{\infty} \frac{ds}{s^{1+\frac{d-1}{2}}}
\exp\Big(-s (M_{j}^{2}(\phi) + 4 \pi^{2} n^{2} T^{2} )-
\frac{\vec{{l}}^{2} L^{2}}{4 s}\Big)\,.
\end{equation}
The vector $\vec{l}$ is defined as $\vec{{l}}=\{ l_{1},l_{2},l_{3}\}$,
and the prime indicates that the term with $\vec{{l}}=0$ is excluded
from the summation. The regulator $\Lambda$ can
also be removed here. Taking the second derivative with respect to
the fields $\phi$ and evaluating at $\phi=0$, we obtain the corresponding
mass correction
\begin{eqnarray}
\delta m_{2}^{2}(T,L) && = 
 \left.\frac{\partial^{2}}{\partial\phi^{2}}
 U_{2}^{\mathrm{{1L}}}(T,L)\right|_{\phi=0}\\ 
 & & \hspace*{-1cm} = \frac{(N+2)\lambda
 TL^{-\frac{d-3}{2}}}{(2\pi)^{\frac{d-1}{2}}}
 \sum_{n=-\infty}^{\infty} \sum_{\{ l_{i}\}}\!'
\Big( \frac{m^{2} + 4\pi^{2} n^{2}
 T^{2}}{\vec{{l}}^{2}}\Big)^{\frac{d-3}{4}} K_{\frac{d-3}{2}}
 \Big(\sqrt{\vec{{l}}^{2} ((mL)^{2} + 4\pi^{2} n^{2} (TL)^{2})}\Big)
 \,.\nonumber  
\end{eqnarray}
We used again the integral representation of the modified Bessel
functions Eq.~\eqref{eq:bessel}. Finally, we show that this contribution
becomes proportional to the temperature $\frac{T}{L}$ for $TL\gg 1$.
For this purpose, we use the asymptotic expansion of the Bessel-functions
for large arguments, given by
\begin{equation}
K_{\nu}(x)\approx \sqrt{\frac{\pi}{x}}\E^{-x}\quad(x\gg1)\,.
\label{eq:bessel_asym}
\end{equation}
Using Eq.~\eqref{eq:bessel_asym}, the mass correction $\delta m_{2}^{2}(T,L)$
for $TL\gg1$ and $d=4$ reads
\be
\delta m_{2}^{2}(T,L) \approx \frac{3(N+2)\lambda}{8^{\frac{1}{2}}\pi}
\frac{T}{L} \sum_{n=-\infty}^{\infty} \sum_{\{
  l_{i}\}}\!'\frac{1}{\sqrt{\vec{{l}}^{2}}} \exp
\left(-\sqrt{\vec{{l}}^{2} \left((mL)^{2}+4\pi^{2}n^{2}(TL)^{2}\right)}\right),
\ee
which drops exponentially in the limit $TL\rightarrow\infty$ for
non-vanishing thermal Matsubara-modes, whereas the contribution from
the zeroth thermal Matsubara-mode ($n=0$) remains finite and is proportional
to the temperature $T$. For $TL\gg1$, the mass-correction $\delta
m_{2}^{2}(T,L)$ is therefore sub-leading, compared to the contribution
$\delta m_{1}^{2}(T)$. 

\bibliographystyle{apsrev}
\bibliography{tcfv}

\end{document}